_______________________________________________________________

\documentstyle[12pt]{article}
\begin{document}

\begin{center}
{\Large On the Stronger Statement of} \\[2mm]
{\Large Levinson's Theorem for the Dirac Equation}

\vspace{5mm}
Zhong-Qi Ma

\vspace{3mm}
{\footnotesize CCAST (World Laboratory),
P.O.Box 8730, Beijing 100081, P. R. of China}\\
{\footnotesize and}\\
{\footnotesize Institute of High Energy Physics, P.O.Box 918(4),
Beijing 100039, P. R. of China}

\vspace{8mm}
\parbox[t]{14cm}{{\small {\bf Abstract}. Recently a stronger statement
of Levinson's theorem for the Dirac equation was presented, where
the limits of the phase shifts at $E=\pm M$ are related to the
numbers of nodes of radial functions at the same energies, respectively.
However, in this letter we show that this statement has to be modified
because the limits of the phase shifts may be negative for the
Dirac equation.}}

\end{center}

\vspace{6mm}
Levinson's theorem $^{[1]}$ is one of the fundamental theorem in
quantum scattering theory. For the Schr\"{o}dinger equation, it gives a
quantitative relation between the limit of the phase shift at zero
energy $\delta_{\ell}(0)$ and the number of bound states $N_{\ell}$:
$$N_{\ell}~=~ \{1 / \pi\}~\left\{\delta_{\ell}(0)
{}~-~\delta_{\ell}(\infty) \right\}~-~\{1/2\}~\sin^{2}(\delta_{\ell})
\eqno (1) $$

\noindent
where the subscript $\ell$ denotes the angular momentum. As is
well know, there is degeneracy of states for the magnetic quantum
number due to the spherical symmetry.
Usually, this degeneracy is not explicitly expressed in the statement
of Levinson's theorem. In (1) the phase shifts are determined by
comparing them with the phase shift at high energy in order to remove their
indetermination of a multiple of $2\pi$. The indetermination
can also be removed by comparing them with the phase shifts of the
free particle, that may be defined as zero$^{[2,3]}$ so that $\delta_{\ell}
(\infty)$ in (1) can be removed. In fact, the experimental
observation for Levinson's theorem shows the jump of the phase shift
at low energy in terms of changing the potential$^{[4]}$. In this letter we
will use this convention for phase shifts. The second term with the
sine square in (1) stands for the half bound state, that was first shown by
Newton$^{[5]}$.

{}From the Sturm-Liouville theorem, the number of nodes of the radial
function at zero energy is equal to the number of the bound states
for the Schr\"{o}dinger equation with a non-singular spherically
symmetric potential, so that it is related to the phase
shift through Levinson's theorem.

Barth\'{e}l\'{e}my$^{[6]}$ first discuss Levinson's theorem for the Dirac
equation by the generalized Jost function. He stated that Levinson's
theorem is valid for positive and negative energies separately as
in the non-relativistic case. But later this statement was found incorrect
$^{[7,3]}$. The correct statement of Levinson's theorem for the
Dirac equation is $^{[7,3]}$:
$$N_{\kappa}~=~ \{1 / \pi\}~\left\{\delta_{\kappa}(M)
{}~+~\delta_{\kappa}(-M) \right\}~-~\{1/2\}~\left\{\sin^{2}(\delta_{\kappa}(M))
{}~+~\sin^{2}(\delta_{\kappa}(-M))\right\} \eqno (2) $$

\noindent
where $\kappa$ is the standard angular momentum parameter that
denotes both the total angular momentum $j$ and the orbital angular
momentum $\ell$: $\kappa=j+1/2$ when $\ell=j+1/2$, and
$\kappa=-(j+1/2)$ when $\ell=j-1/2$. $N_{\kappa}$ and $\delta_{\kappa}(E)$
are the number of bound states and the phase shift at energy $E$
for the given angular momentum $\kappa$, respectively. The key point of
Levinson's theorem for the Dirac equation is that the limits of the phase
shifts at the thresholds $E=\pm M$ (zero momentum) may be positive, zero
or negative. For example, as the attractive potential becomes strong
enough, a scattering state of positive energy may change to a bound
state, and furthermore, change to a scattering state of negative
energy. In this case the limit $\delta_{\kappa}(-M)$ of phase shift
at $E=-M$ becomes negative. On the other hand, the limit
of the phase shift at zero energy for the Schr\"{o}dinger equation
only can be zero or positive, if the spherically symmetric potential
is continuous and finite.

Recently, Poliatzky$^{[8,9]}$ transformed the Dirac equation into couple
of effective Schr\"{o}dinger-type equations near the thresholds
$E=\pm M$ and showed a stronger statement of Levinson's theorem
for the Dirac equation:
$$n_{\kappa}(\pm M)~=~\delta_{\kappa}(\pm M)/\pi ~-~\{1/2\}~
\sin^{2}(\delta_{\kappa}(\pm M)) \eqno (3) $$

\noindent
where $n_{\kappa}(\pm M)$ denote the numbers of nodes for the radial
functions with energies $\pm M$, respectively$^{[11]}$. In Ref.[8]
$n_{\kappa}(\pm M)$ are explained as the numbers of bound states of
the effective Schr\"{o}dinger-like equations, that are not easy to
count. For the potential $|V(r)|<2M$, two numbers are equal to each
other due to the Sturm-Liouville theorem.

A few viewpoints in [8] and [9] have drawn some discussions. Newton$^{[10]}$
criticized the modification of the usual quasi-orthogonality of the
scattering wave function. Poliatzky's explanation$^{[11]}$ that "it is
merely a short-hand description of two equations" is acceptable.

Newton's another criticism is that the limit of phase shift at the
threshold cannot be $\pi/4$ in addition to a multiple of $\pi$.
In the reply paper$^{[11]}$ Poliatzky agreed that there is a certain
constraint to rule out this surprising case$^{[12]}$. However, he still
stated that this case cannot be ruled out completely in a conventional
proof of Levinson's theorem. As a matter of fact, in the proof of
Levinson's theorem by the Sturm-Liouville theorem the surprising case
has been ruled out completely. (see (15b) in Ref.[2] and (14) in Ref.[3])

It is easy to see that the stronger statement (3) of Levinson's theorem
for the Dirac equation is not generally correct. The main point is
that the limit of the phase shifts at thresholds for the Dirac equation
may be negative, but the number of nodes, or the number of bound
states, is a non-negative integer.
Why may the limit of the phase shift be negative for an effective
Schr\"{o}dinger-type equation? The reason is that the effective
potential in the effective Schr\"{o}dinger-type equations
is singular when the potential $V$ is finite, but strong
enough (see Eqs.(22) and (23) in [8] when $V=\pm 2M$). For those
equations Levinson's theorem is not effective. In the same reason,
Poliatzky's proof for the statement that the sum of the numbers of nodes,
$n_{\kappa}(M)+n_{\kappa}(-M)$, is equal to the number of bound
states $N_{\kappa}$ is wrong.
Besides, there are two radial functions for the wave function of the
Dirac equation. The numbers of two radial functions may not be same.
One has to determine which radial function the nodes
counted in (3) belong to. It seems that Poliatzky assumed$^{[8]}$ that
$n_{\kappa}(M)$ is the number of nodes of $u_{1M\kappa}$, and
$n_{\kappa}(-M)$ is that of $u_{2-M\kappa}$.

Now, we are going to study this problem in more detail. Discuss the
Dirac equation with a spherically symmetric potential $V(r)$.
For simplicity, we assume the potential is cutoff: $V(r)=0$, if
$r>r_{0}$. Let
$$\Psi_{\kappa mE}(r)~=~\displaystyle {1\over r}\left( \begin{array}{c}
if_{\kappa E}(r)\phi_{\kappa m}(\theta,\varphi) \\
g_{\kappa E}(r)\phi_{-\kappa m}(\theta,\varphi) \end{array} \right)
\eqno (4) $$

\noindent
where $\phi_{\kappa m}(\theta,\varphi)$ is the two-component spherical
spinor$^{[7]}$ and the radial functions $f$ and $g$ satisfy the radial
equation:
$$\begin{array}{l}
f'_{\kappa E}~+~\displaystyle {\kappa \over r}f_{\kappa E}~=~-~
(E-V+M)g_{\kappa E},\\[2mm]
g'_{\kappa E}~-~\displaystyle {\kappa \over r}g_{\kappa E}~=~
(E-V-M)f_{\kappa E}
\end{array} \eqno (5) $$

\noindent
where the radial functions $f$ and $g$ are proportional to $u_{1}$
and $-u_{2}$ used in [8], respectively.

The solution with $-\kappa$ and $E$ can be obtained from that with
$\kappa$ and $-E$ by exchange $f_{-\kappa E}\leftrightarrow
g_{\kappa -E}$ and by the replacement $V\rightarrow -V$.
In the following we only discuss the solution with positive $\kappa$.
Assume $\kappa>1$ if without notification. Near the origin there is
only one physically admissible solution so that the ratio of two
radial functions at $r=r_{0}$ is determined.

In the region $r>r_{0}$ we have $V=0$.
There are two oscillatory solutions when $|E|>M$. Through appropriate
combination they can meet the match condition at $r=r_{0}$:
$$\left. \displaystyle {f_{\kappa E}(r)\over g_{\kappa E}(r)}
\right|_{r_{0}-}~\equiv ~A_{\kappa}(E)~=~
\left. \displaystyle {f_{\kappa E}(r)\over g_{\kappa E}(r)}
\right|_{r_{0}+} \eqno (6) $$

\noindent
The phase shift is determined from this condition (see (14) in [3]):
$$\tan \delta_{\kappa}(E)~\sim~-~\displaystyle {\pi (kr_{0})^{2\kappa -1}
\over (2\kappa -1)!!(2\kappa -3)!!}~{A_{\kappa}(E)+2Mr_{0}/(2\kappa+1)
\over A_{\kappa}(E)+2M(2\kappa -1)/(k^{2}r_{0}) }, \eqno (7a)$$

\noindent
when $E>M$ and $E\sim M$, and
$$\tan \delta_{\kappa}(E)~\sim~-~\displaystyle {\pi (kr_{0})^{2\kappa -1}
\over (2\kappa -1)!!(2\kappa -3)!!}~{A_{\kappa}(E)-
k^{2}r_{0}/\{2M(2\kappa+1)\}
\over A_{\kappa}(E)-\displaystyle {2\kappa -1 \over 2Mr_{0}}\left[
1-\displaystyle {k^{2}r_{0}^{2} \over (2\kappa -1)(2\kappa -3) }\right]},
\eqno (7b) $$

\noindent
when $E<-M$ and $E\sim -M$, where $k=(E^{2}-M^{2})^{1/2}$. Therefore,
the phase shift is monotonic with respect to the ratio
$A_{\kappa}(E)$ (see (15) in [3]):
$$\left. \displaystyle {\partial \delta_{\kappa}(E) \over
\partial A_{\kappa}(E)} \right|_{E} <0 ~~{\rm when}~~E>M,~~~~{\rm and}~~~~
\left. \displaystyle {\partial \delta_{\kappa}(E) \over
\partial A_{\kappa}(E)} \right|_{E} >0 ~~{\rm when}~~E<-M
 \eqno (8) $$

\noindent
{}From (7) we see that the limits of the phase shifts at $E=\pm M$ are
multiples of $\pi$ when $\kappa>1$. When $A_{\kappa}(M)$
decreases (or increases) across infinity ($g_{\kappa M}(r_{0})=0$),
$\delta_{\kappa}(M)$ jumps by a $\pi$ (or $-\pi$), and when
$A_{\kappa}(-M)$ decreases (or increases) across the value $\rho_{1}$,
$\rho_{1}=(2\kappa -1)/(2Mr_{0})$,
$\delta_{\kappa}(-M)$ jumps by a $-\pi$ (or $\pi$). When $\kappa=1$
and $A_{1}(-M)$ decreases (or increases) to the value $\rho_{1}$,
$\delta_{1}(-M)$ jumps by a $-\pi/2$ (or $\pi/2$). For this critical
case where $A_{1}(-M)=\rho_{1}$, there is a finite solution at $E=-M$, called
half bound state (see (16) for $\kappa=1$). The case that
$\delta_{\kappa}(M)$ or $\delta_{\kappa}(-M)$ is equal to $\pi/4$
in addition to a multiple of $\pi$ is ruled out by (7).

On the other hand, when $|E|<M$ there is only one convergent
solution in the region $r>r_{0}$ so that the match condition (6)
is not always satisfied. When the condition (6) is satisfied,
a bound state appears at this energy. The generalized Sturm-Liouville
theorem says (see (17) in [3]) that as the energy $E$ or $-V$
increases, the ratio $A_{\kappa}(E)$ decreases monotonically, and as the
energy $E$ increases, the ratio
$f_{\kappa E}(r_{0}+)/g_{\kappa E}(r_{0}+)$ with $|E|<M$
increases monotonically. From the solution in the region $r>r_{0}$
we have (see (18b) in [3]):
$$\left. \displaystyle {f_{\kappa E}(r) \over
g_{\kappa E}(r)} \right|_{r_{0}+}~=~\left\{\begin{array}{ll}
\infty &{\rm if}~~ E=M\\ \displaystyle {2\kappa -1 \over 2Mr_{0}}~
\equiv~\rho_{1}~~&{\rm if}~~E=-M \end{array} \right. \eqno (9) $$

\noindent
If $V=0$, we have (see (18a) in [3]):
$$A_{\kappa}(M)~=~\left\{\begin{array}{ll} - \displaystyle {2Mr_{0}
\over 2\kappa +1}~\equiv~-\rho_{2}~~&{\rm if}~~E=M\\ 0 &{\rm of}~~
E=-M \end{array} \right. \eqno (10) $$

It is easy to see from (9) and (10) that there is no overlap
for two varying ranges of two ratios such that there is no bound state
for the case $V=0$. When $V$ decreases from zero,
$f_{\kappa E}(r_{0}+)/g_{\kappa E}(r_{0}+)$ with $|E|<M$ does not
change, but $A_{\kappa}(E)$ does decrease. Each time $A_{\kappa}(M)$
decreases across infinity ($g_{\kappa E}(r_{0})=0$), a new overlap for
two ranges occurs, namely, a scattering state of
positive energy becomes a bound state. At the same time
$\delta_{\kappa}(M)$ jumps by $\pi$. On the other hand,
each time $A_{\kappa}(-M)$ decreases across the value $\rho_{1}$,
an overlap disappears, so that a bound state becomes a scattering
state of negative energy, $\delta_{\kappa}(-M)$ jumps by $-\pi$.
If $V$ increases from zero (repulsive potential), the process
goes conversely, namely, if $A_{\kappa}(-M)$ increases across the
value $\rho_{1}$, a scattering state of negative energy changes to a
bound state and $\delta_{\kappa}(-M)$ jumps by $\pi$, and if
$A_{\kappa}(M)$ increases across infinity, a bound state changes
to a scattering state of positive energy and $\delta_{\kappa}(M)$ jumps
by $-\pi$. For $\kappa=1$ we have to consider the possible half
bound state.

It is the essence of Levinson's theorem for the Dirac equation.

Now, we give a calculable example to show the relations between
the phase shifts and the numbers of nodes explicitly. Consider
a square well potential:
$$V(r)~=~\left\{\begin{array}{ll}-~\lambda ~~&{\rm if}~~r\leq r_{0}\\
0 &{\rm if}~~r>r_{0} \end{array} \right. \eqno (11) $$

\noindent
The potential $V$ is attractive when $\lambda>0$, and is repulsive
when $\lambda<0$. Solve (5) at the energies $E=\pm M$.

i) $E=M$

In the region $r_{0}<r$ where $V=0$, the solution for $E=M$ is:
$$\begin{array}{l}
f_{\kappa M}(r)~=~f_{\kappa M}(r_{0})~\left(\displaystyle {r \over r_{0}
}\right)^{-\kappa}~-~\displaystyle {2Mr_{0} \over 2\kappa +1}~
g_{\kappa M}(r_{0})~\left\{\left(\displaystyle {r \over r_{0}
}\right)^{\kappa+1}~-~\left(\displaystyle {r \over r_{0}
}\right)^{-\kappa}\right\}\\
g_{\kappa M}(r)~=~g_{\kappa M}(r_{0})~\left(\displaystyle {r \over r_{0}
}\right)^{\kappa}
\end{array} \eqno (12) $$

\noindent
In this region there is no node for $g_{\kappa M}(r)$, unless
$g_{\kappa M}(r_{0})=0$. In the latter case $g_{\kappa M}(r)=0$
for $r\geq r_{0}$. For this case we say $g_{\kappa M}(r)$
has a node at $r=r_{0}$ as usual. $f_{\kappa M}(r)$ has one node
in the region $r>r_{0}$ if $A_{\kappa}(M)>0$, and no node if
$A_{\kappa}(M)\leq 0$. Obviously, $f_{\kappa M}(r)$ has a node at
$r_{0}$ when $A_{\kappa}(M)=0$. When $A_{\kappa}(M)$ goes to infinity,
($g_{\kappa M}(r_{0})=0$) we have a bound state of energy $M$:
$$\begin{array}{l}
f_{\kappa M}(r)~=~f_{\kappa M}(r_{0})~\left(\displaystyle {r \over r_{0}
}\right)^{-\kappa},~~~~
g_{\kappa M}(r)~=~0,~~~~{\rm for}~~r\geq r_{0}
\end{array} \eqno (13) $$

{}From the solutions for $E=M$ in the region $0\leq r \leq r_{0}$ $^{[3]}$,
we obtain the ratio $A_{\kappa}(M)$ that decreases monotonically
as $\lambda$ increases:
$$A_{\kappa}(M)~=~\left\{\begin{array}{ll}
-\left\{\displaystyle {2M+\lambda \over \lambda}\right\}^{1/2}
{}~\displaystyle {J_{\kappa+1/2}(p_{1}r_{0}) \over
J_{\kappa-1/2}(p_{1}r_{0})},~~~~{\rm if}~~\lambda \geq 0\\
\left\{\displaystyle {2M-|\lambda| \over |\lambda|}\right\}^{1/2}
{}~\displaystyle {iJ_{\kappa+1/2}(ip_{2}r_{0}) \over
J_{\kappa-1/2}(ip_{2}r_{0})},~~~~{\rm if}~~0 > \lambda > -2M\\
\left\{\displaystyle {|\lambda|-2M \over |\lambda|}\right\}^{1/2}
{}~\displaystyle {J_{\kappa+1/2}(p_{3}r_{0}) \over
J_{\kappa-1/2}(p_{3}r_{0})},~~~~{\rm if}~~\lambda \leq -2M
\end{array} \right. \eqno (14) $$

\noindent
where $p_{1}=\{|\lambda|(|\lambda|+2M)\}^{1/2}$,
$p_{2}=\{|\lambda|(2M-|\lambda|)\}^{1/2}$ and
$p_{3}=\{|\lambda|(|\lambda|-2M)\}^{1/2}$, and
$J_{\ell}(r)$ is the Bessel function.

ii) $E=-M$

The solution in the region $r_{0}<r$ for $E=-M$ is:
$$\begin{array}{l}
f_{\kappa -M}(r)~=~f_{\kappa -M}(r_{0})~\left(\displaystyle {r \over r_{0}
}\right)^{-\kappa}\\
g_{\kappa -M}(r)~=~g_{\kappa -M}(r_{0})~\left(\displaystyle {r \over r_{0}
}\right)^{\kappa}~+~\displaystyle {2Mr_{0} \over 2\kappa -1}~
f_{\kappa -M}(r_{0})~\left\{\left(\displaystyle {r \over r_{0}
}\right)^{-\kappa+1}~-~\left(\displaystyle {r \over r_{0}
}\right)^{\kappa}\right\}
\end{array} \eqno (15) $$

\noindent
In this region there is always a node at infinity for $f_{\kappa -M}(r)$.
When $f_{\kappa -M}(r_{0})=0$, $f_{\kappa -M}(r)=0$ in the region
$r\geq r_{0}$. $g_{\kappa -M}(r)$ has no node if $A_{\kappa}(-M)<\rho_{1}$
and one node if $A_{\kappa}(-M)\geq \rho_{1}$. When $A_{\kappa}(-M)
=\rho_{1}$ we have a bound state of energy $-M$:
$$f_{\kappa -M}(r)~=~f_{\kappa -M}(r_{0})~\left(\displaystyle {r \over r_{0}
}\right)^{-\kappa},~~~~
g_{\kappa -M}(r)~=~g_{\kappa -M}(r_{0})~
\left(\displaystyle {r \over r_{0}}\right)^{-\kappa+1}
 \eqno (16) $$

\noindent
When $\kappa=1$ it is a half bound state. From the solutions for $E=-M$
in the region $r\leq r_{0}$$^{[3]}$ we obtain the ratio
$A_{\kappa}(-M)$ that decreases monotonically as $\lambda$ increases:
$$A_{\kappa}(-M)~=~\left\{\begin{array}{ll}
\left\{\displaystyle {|\lambda| \over |\lambda|+2M}\right\}^{1/2}
{}~\displaystyle {J_{\kappa+1/2}(p_{1}r_{0}) \over
J_{\kappa-1/2}(p_{1}r_{0})},~~~~{\rm if}~~\lambda \leq 0\\
-\left\{\displaystyle {\lambda \over 2M-\lambda}\right\}^{1/2}
{}~\displaystyle {iJ_{\kappa+1/2}(ip_{2}r_{0}) \over
J_{\kappa-1/2}(ip_{2}r_{0})},~~~~{\rm if}~~0 < \lambda < 2M\\
-\left\{\displaystyle {\lambda \over \lambda-2M}\right\}^{1/2}
{}~\displaystyle {J_{\kappa+1/2}(p_{3}r_{0}) \over
J_{\kappa-1/2}(p_{3}r_{0})},~~~~{\rm if}~~\lambda \geq 2M
\end{array} \right. \eqno (17) $$

Carefully counting the nodes of $f_{\kappa \pm M}(r)$ and
$g_{\kappa \pm M}(r)$ when $\lambda$ increases and decreases from
zero, we reach the following conclusions:

a) (2) is the correct statement of Levinson's theorem for the
Dirac equation.

b) As a stronger statement of Levinson's theorem for the Dirac
equation, (3) is not general correct, especially for a very strong
potential ($|V| \geq 2M$).

c) In the example of a square well potential, (3) can be modified
as follows. For positive $\kappa$, if $\delta_{\kappa}(M)\geq 0$,
$\delta_{\kappa}(M)/\pi$ is equal to the number of nodes of
$g_{\kappa M}(r)$, which is the same as that of $f_{\kappa M}(r)$;
if $\delta_{\kappa}(M)< 0$, $-\delta_{\kappa}(M)/\pi$
is equal to the number of nodes of $g_{\kappa M}(r)$ in the region
$0<r<r_{0}$; if $\delta_{\kappa}(-M)\geq 0$, $(\delta_{\kappa}(-M)/\pi)
-\sin^{2}(\delta_{\kappa}(-M))/2$
is equal to the number of nodes of $g_{\kappa -M}(r)$; and if
$\delta_{\kappa}(-M)< 0$, $-(\delta_{\kappa}(-M)/\pi)+\sin^{2}(
\delta_{\kappa}(-M))/2$
is equal to the number of nodes of $g_{\kappa -M}(r)$ in the region
$0<r<r_{0}$ subtracting the number of its nodes in the region $r>r_{0}$.
However, it is still an open problem whether there is and what is
the stronger statement of Levinson's theorem for the Dirac equation.

\vspace{5mm}
{\bf Acknowledgments}. The author would like to thank Professor
Chen Ning Yang for drawing his attention to the papers
[8] and [9]. This work was supported by the National
Natural Science Foundation of China and Grant No. LWTZ-1298 of
Chinese Academy of Sciences.

\end{document}